# The Strong Sensitivity of the Characteristics of Binary Stochastic Neurons Employing Low Barrier Nanomagnets to Small Geometrical Variations

Rahnuma Rahman, and Supriyo Bandyopadhyay, *Fellow, IEEE*

*Abstract*—Binary stochastic neurons (BSNs) are excellent activators for machine learning. An ideal platform for implementing them are low- or zero-energy-barrier nanomagnets (LBMs) possessing in-plane anisotropy (e.g. circular or slightly elliptical disks) whose fluctuating magnetization encodes a probabilistic (p-) bit. Here, we show that such a BSN's activation function, the pinning current (which pins the output to a particular binary state), and the response time – all exhibit strong sensitivity to very slight geometric variations in the LBM's cross-section. A mere 1% change in the diameter of a circular nanomagnet in any arbitrary direction can alter the response time by a factor of ~4 at room temperature and a 10% change can alter the pinning current by a factor of ~2. All this causes large device-to-device variation which is detrimental to integration.

We also show that the energy dissipation is lowered but the response time is increased by replacing a circular cross-section with a slightly elliptical one and then encoding the p-bit in the magnetization component along the major axis. Encoding the p-bit in the magnetization component along the minor axis has the opposite effect. The energy-delay-product, however, is relatively independent of whether the cross-section is a circle or an ellipse and which magnetization component encodes the p-bit in the case of the ellipse.

*Index Terms*—Binary stochastic neurons, low barrier nanomagnets, pinning current, correlation time, device-to-device variation

## I. Introduction

ARTIFICIAL intelligence (AI) platforms typically employ deep neural networks (DNNs). DNNs are also dominant in statistical machine learning which includes advanced computer vision (image detection, recognition and segmentation [1]), natural language processing [2], etc. There is now a strong desire to improve the energy and speed performances of these systems, leading to the interest in spiking neural networks (SNN). One class of SNNs can be activated by binary stochastic impulses [2-4] and can be implemented with binary stochastic neurons (BSN) that have two distinct output states (-1 and +1). A BSN will output either a -1 or a +1 (in the bipolar representation) with a probability determined by a specific function of the input that is provided to it. It is also an excellent representation for Ising machines used to solve combinatorial optimization problems [5].

Low (or zero) energy barrier nanomagnets (LBMs) possessing in-plane magnetic anisotropy are a natural (and popular) choice for BSNs and have been shown to be capable of acting as efficient hardware accelerators for machine learning [5-7]. The magnetization vectors of these LBMs, which are circular or nearly circular disks, fluctuate at room temperature owing to thermal perturbations and the fluctuating (stochastic) magnetization can produce, with suitable design, either a -1 or a +1 state [6, 7] whose probabilities are engineered by passing a spin polarized current through the LBM [6, 7]. The current, which is the input to the BSN, biases the output towards either -1 or +1 and the degree of bias can be tuned with the current. At any time step $n+1$, the BSN will output a state given by the analytical expression

$$m_i(n+1) = \text{sgn}\left[\tanh(I_i(n)) - r_i\right] \quad (1)$$

where $I_i$ is the dimensionless input spin current that biases the output either towards 0 or towards 1 (depending on its sign) and $r_i$ is a random number uniformly distributed between -1 and +1. Each BSN described by Equation (1) receives its input from a weighted sum of other BSNs obtained from a synapse $I_i(n) = \sum_j W_{ij} m_j(n)$. A wide variety of problems can be solved by properly designing or learning the weights $W_{ij}$, e.g. classification problems [8], constrain satisfaction problems [9], generation of cursive letters [10], etc.

There are three critical parameters for a BSN. The first is the "activation function" which represents the probability of outputting a given state (-1 or +1) as a function of the input. In the case of LBMs implementing BSNs, the input is a spin polarized current of magnitude $I_0$ and the activation function is the "$\langle \vec{s} \bullet \vec{m} \rangle$ versus $I_0$" relation where $\vec{s}$ and $\vec{m}$ are the unit vectors in the direction of the spin polarization of the current and the magnetization of the LBM, respectively [6, 7]. The angular bracket denotes time averaging. The quantity $\vec{s} \bullet \vec{m}$ is

This work was supported by the US National Science Foundation under grants CCF-2001255 and CCF-2006843.

Rahnuma Rahman and Supriyo Bandyopadhyay are with the Department of Electrical and Computer Engineering, Virginia Commonwealth University, Richmond, VA 23284, USA

Color versions of one or more of the figures in this article are available online at http://ieeexplore.ieee.org.

the component of the magnetization along the direction of the current's spin polarization at any instant of time. A positive value will represent the bit 1 and a negative value the bit -1. If $p$ is the time averaged probability of outputting bit 1 for a given $I_0$, then $p = \left[\langle \vec{s} \bullet \vec{m} \rangle + 1\right]/2$. The second critical parameter is the "pinning current" which is the minimum value of the current for which $p = 1$, The third critical parameter, which determines the response speed of BSNs implemented with LBMs, is the "correlation time" $\tau_c$ which is the full-width-at-half maximum (FWHM) of the decay characteristic of the auto-correlation function of the magnetization fluctuations [6]. It is also called the "response time". LBMs with in-plane anisotropy typically have a value of $\tau_c$ that is about two orders of magnitude smaller than LBMs with perpendicular magnetic anisotropy [6] and are therefore favored because the BSN speed (i. e. the speed of hardware accelerators) will be higher if they are employed instead of ones with perpendicular anisotropy.

In this paper, we present *four* important results pertaining to BSNs implemented with LBMs possessing in-plane anisotropy. First, we show that the activation function ($\langle \vec{s} \bullet \vec{m} \rangle$ vs $I_0$) is quite sensitive to small changes in the shape of the nanomagnet's cross section. This poses a serious problem since the activation function determines what the probability of producing a +1 output is for a given current. This probability must be controllable by the current and that controllability is lost if the activation function exhibits significant device to device variability due to the strong sensitivity to small shape variations.

Second, we show that the pinning current, which is the current required to pin the output to either +1 or -1 with 100% probability also depends on the cross-sectional shape.

Third, we show that the auto-correlation function of the magnetization fluctuation and its FWHM (i. e. $\tau_c$) can change significantly depending on whether the LBM is perfectly circular or slightly elliptical. We show this with the example of a perfectly circular LBM of diameter 100 nm and a very slightly elliptical LBM with major axis of 100 nm and minor axis of 99 nm. Their $\tau_c$-s differ by a factor of 3-4, even though their lateral dimensions in just one direction differ by a mere 1%. The circular nanomagnet has the longer (shorter) correlation time if $\vec{s}$ is along the minor (major) axis of the elliptical nanomagnet. The quantity $\tau_c$ is proportional to the "response time" of the BSN. The significance of the response time is that a BSN does not begin to produce an output of +1 with a probability $p$ (and -1 with a probability 1-$p$) as soon as the bias current is set for $p$. It takes a certain amount of time for that to happen, which is the response time. If this time varies significantly, then the circuits will become error-prone since one would not know exactly how long the output has to be sampled to obtain correct results.

Fourth, we show that "spin-inertia" [11], which causes short-lived nutation in magneto-dynamics and can sometimes have long-term effects [12], has no significant effect on the auto-correlation decay characteristics of BSNs implemented with LBMs and hence has no effect on $\tau_c$. It also has no perceptible effect on the activation function. That is reassuring since spin inertia cannot be easily controlled or engineered.

## II. THEORY

To study the dynamical characteristics of BSNs implemented with LBMs, such as a circular or slightly elliptical disk shown in Fig. 1, we first simulate their magneto-dynamics in the presence of thermal noise by solving the stochastic Landau-Lifshitz-Gilbert (LLG) equation (sometimes referred to as the Landau-Lifshitz-Gilbert-Langevin equation) with the effect of spin-inertia included as a second order time derivative of the magnetization with a characteristic pre-factor $\tau$ which is the time scale over which the angular momentum of magnetization relaxes [11]. It is also the time scale over which nutation takes place in the magneto-dynamics [11]. Additionally, we model the effect of a spin polarized dc current flowing normal to the cross-section of the nanomagnet with the spin polarized along either the *x*-axis or the *y*-axis (minor and major axes in the case of an elliptical nanomagnet) shown in Fig. 1. This allows us to evaluate the activation function and the pinning current which will pin the magnetization along the direction of spin polarization which is equivalent to pinning it to either the state +1 or the state -1 if we assume that magnetization in the direction of spin polarization encodes +1 and in the opposite direction encodes -1. Pinning the magnetization to the minor axis (hard axis) with a current will be harder than pinning it to the major axis (easy axis) since the easy axis orientation corresponds to the potential energy minimum and the hard axis orientation corresponds to the potential energy maximum in an elliptical nanomagnet. Therefore, the pinning current will be larger in the former case.

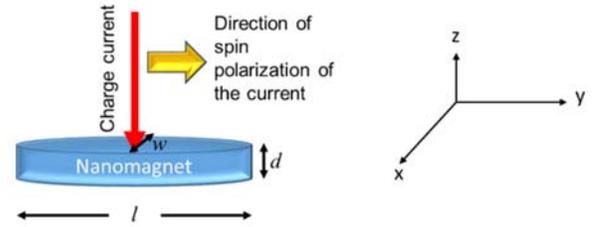

**Fig. 1**: A nanomagnet shaped like an elliptical disk with the major axis along the *y*-axis and minor axis along the *x*-axis.

The stochastic LLG equation including the effect of spin inertia is [12]

$$\frac{d\vec{m}(t)}{dt} = -|\gamma|\vec{m}(t) \times \left[\vec{H}(t) - \frac{\alpha}{|\gamma|}\left(\frac{d\vec{m}(t)}{dt} + \tau\frac{d^2\vec{m}(t)}{dt^2}\right)\right] \\ + a\vec{m}(t) \times \left[\frac{\xi \vec{I}_s \mu_B}{qM_s\Omega} \times \vec{m}(t)\right] + b\frac{\xi \vec{I}_s \mu_B}{qM_s\Omega} \times \vec{m}(t) \quad (2)$$

where $\vec{H}(t) = \vec{H}_{thermal}(t) + \vec{H}_{demag}(t)$

$$\vec{H}_{thermal}(t) = \sqrt{\frac{2\alpha kT}{\gamma(1+\alpha^2)\mu_0 M_s \Omega \Delta t}}$$
$$\times \left[ G^x_{(0,1)}(t)\hat{x} + G^y_{(0,1)}(t)\hat{y} + G^z_{(0,1)}(t)\hat{z} \right]$$

$$\vec{H}_{demag}(t) = -M_s N_{d-xx} m_x(t)\hat{x} - M_s N_{d-yy} m_y(t)\hat{y} - M_s N_{d-zz} m_x(t)\hat{z}$$

and

$\vec{I}_s = I_0 \hat{x}$ (for spin polarized in the *x*-direction)
$\vec{I}_s = I_0 \hat{y}$ (for spin polarized in the *y*-direction)

The demagnetization factors are given by [15]

$$N_{d-xx} = \frac{\pi}{4}\left(\frac{d}{l}\right)\left[1 + \frac{5}{4}\left(\frac{l-w}{l}\right) + \frac{21}{16}\left(\frac{l-w}{l}\right)^2\right]$$

$$N_{d-yy} = \frac{\pi}{4}\left(\frac{d}{l}\right)\left[1 - \frac{1}{4}\left(\frac{l-w}{w}\right) - \frac{3}{16}\left(\frac{l-w}{l}\right)^2\right]$$

$$N_{d-zz} = 1 - N_{d-xx} - N_{d-yy}$$

Here $\vec{m}(t)$ is the magnetization vector normalized to the saturation magnetization of the magnetic material $M_s$, and $m_i(t)$ is its *i*-th scalar component. The quantity $\vec{H}(t) = \vec{H}_{demag} + \vec{H}_{thermal}$ where the first term is due to any shape anisotropy of the LBM and the second is due to thermal noise and is random. The magnetization vector precesses around $\vec{H}(t)$ and hence fluctuates randomly.

In Equation (2), the quantity $\gamma$ is the gyromagnetic precession constant ($2.21\times10^5$ rad-m/A-s), $\alpha$ is the Gilbert damping factor in the nanomagnet associated with damping of the precession (a material constant), and $\tau$ is the relaxation time of the angular momentum associated with spin inertia (the timescale over which nutation takes place [11]). The magnitude of the charge current injected into the nanomagnet is $I_0$ and the degree of spin polarization of the current is $\xi$. The spin in the current is polarized along the intended pinning direction, which is either the major axis of the nanomagnet (*y*-direction) or the minor axis (*x*-direction) for the ellipse in this study.

The last two terms in Equation (2) represent the Slonczewski and field-like torques, respectively, with $\mu_B$ being the Bohr magneton. The relative strengths of these two torques are given by the quantities *a* and *b*, and following ref. [13], we assume $a = 1.0$ and $b = 0.3$. The value of $\tau$ can range from a few fs to ~100 ps in ferromagnets [14].

The demagnetizing field $\vec{H}_{demag}(t)$ depends on the demagnetization coefficients $N_{d-xx}$, $N_{d-yy}$ and $N_{d-zz}$, which, in turn, depend on the nanomagnet dimensions $l$ (major axis), $w$ (minor axis) and $d$ (thickness) as shown in Fig. 1. In the expression for the thermal noise field term $\vec{H}_{thermal}(t)$, $\Omega$ is the volume of the nanomagnet $\left[\Omega = (\pi/4)l \times w \times d\right]$, $\Delta t$ is the time step used in the simulation, $\mu_0$ is the permeability of free space and the quantities $G^i_{(0,1)}$ ($i = x, y, z$) are three statistically independent Gaussians with zero mean and unit standard deviation [16]. All simulations assume room temperature ($T = 300$ K). We also assume the following material parameters: $M_s = 8\times10^5$ A/m, $\alpha = 0.1$. This corresponds to a material like cobalt or Terfenol-D. The time step chosen is $\Delta t = 0.1$ ps and it was found in numerous previous simulations (e.g. [12, 13] and references therein) that making the time step any smaller does not make any perceptible change (1% or more) in the magneto-dynamics results.

The vector equation (2) is decomposed into three coupled scalar equations. When the spin in the injected current is polarized along the *x*-direction, the coupled equations are:

$$\frac{dm_x(t)}{dt} = -|\gamma|\left[m_y(t)H_z(t) - m_z(t)H_y(t)\right]$$
$$+\alpha\left[m_y(t)\frac{dm_z(t)}{dt} - m_z(t)\frac{dm_y(t)}{dt}\right]$$
$$+\alpha\left[m_y(t)\tau\frac{dm_z^2(t)}{dt^2} - m_z(t)\tau\frac{dm_y^2(t)}{dt^2}\right]$$
$$+a\left(\frac{\xi I_0 \mu_B}{qM_s\Omega}\right)\left[m_y^2(t) + m_z^2(t)\right]$$

$$\frac{dm_y(t)}{dt} = -|\gamma|\left[m_z(t)H_x(t) - m_x(t)H_z(t)\right]$$
$$+\alpha\left[m_z(t)\frac{dm_x(t)}{dt} - m_x(t)\frac{dm_z(t)}{dt}\right]$$
$$+\alpha\left[m_z(t)\tau\frac{dm_x^2(t)}{dt^2} - m_x(t)\tau\frac{dm_z^2(t)}{dt^2}\right]$$
$$-b\left(\frac{\xi I_0\mu_B}{qM_s\Omega}\right)m_z(t) - a\left(\frac{\xi I_0\mu_B}{qM_s\Omega}\right)m_x(t)m_y(t)$$

$$\frac{dm_z(t)}{dt} = -|\gamma|\left[m_x(t)H_y(t) - m_y(t)H_x(t)\right]$$
$$+\alpha\left[m_x(t)\frac{dm_y(t)}{dt} - m_y(t)\frac{dm_x(t)}{dt}\right] \quad (3a)$$
$$+\alpha\left[m_x(t)\tau\frac{dm_y^2(t)}{dt^2} - m_y(t)\tau\frac{dm_x^2(t)}{dt^2}\right]$$
$$+b\left(\frac{\xi I_0\mu_B}{qM_s\Omega}\right)m_y(t) - a\left(\frac{\xi I_0\mu_B}{qM_s\Omega}\right)m_z(t)m_x(t)$$

where $H_i(t)$ ($i = x, y, z$) is the *i*-th component of the effective magnetic field at time *t*.

When the spin in the injected current is polarized in the *y*-direction, the coupled equations change to

$$\frac{dm_x(t)}{dt} = -|\gamma|\left[m_y(t)H_z(t) - m_z(t)H_y(t)\right]$$
$$+\alpha\left[m_y(t)\frac{dm_z(t)}{dt} - m_z(t)\frac{dm_y(t)}{dt}\right]$$
$$+\alpha\left[m_y(t)\tau\frac{dm_z^2(t)}{dt^2} - m_z(t)\tau\frac{dm_y^2(t)}{dt^2}\right]$$
$$+b\left(\frac{\xi I_0 \mu_B}{qM_s\Omega}\right)m_z(t) - a\left(\frac{\xi I_0 \mu_B}{qM_s\Omega}\right)m_x(t)m_y(t)$$

$$\frac{dm_y(t)}{dt} = -|\gamma|\left[m_z(t)H_x(t) - m_x(t)H_z(t)\right]$$
$$+\alpha\left[m_z(t)\frac{dm_x(t)}{dt} - m_x(t)\frac{dm_z(t)}{dt}\right]$$
$$+\alpha\left[m_z(t)\tau\frac{dm_x^2(t)}{dt^2} - m_x(t)\tau\frac{dm_z^2(t)}{dt^2}\right]$$
$$+a\left(\frac{\xi I_0 \mu_B}{qM_s\Omega}\right)\left[m_x^2(t) + m_z^2(t)\right]$$

$$\frac{dm_z(t)}{dt} = -|\gamma|\left[m_x(t)H_y(t) - m_y(t)H_x(t)\right]$$
$$+\alpha\left[m_x(t)\frac{dm_y(t)}{dt} - m_y(t)\frac{dm_x(t)}{dt}\right] \quad (3b)$$
$$+\alpha\left[m_x(t)\tau\frac{dm_y^2(t)}{dt^2} - m_y(t)\tau\frac{dm_x^2(t)}{dt^2}\right]$$
$$-b\left(\frac{\xi I_0 \mu_B}{qM_s\Omega}\right)m_x(t) - a\left(\frac{\xi I_0 \mu_B}{qM_s\Omega}\right)m_z(t)m_y(t)$$

These three coupled equations [3(a) or 3(b)] are solved using MATLAB to find the magnetization components $m_x(t), m_y(t), m_z(t)$ at any time $t$.

### III. SIMULATION RESULTS

#### 3.1 Pinning current and activation function of the BSN

Equation (2) contains a random term associated with the thermal noise field $\vec{H}_{thermal}$ and hence the magnetization versus time trajectories $[m_i(t) \text{ vs } t]; i = x, y, z$ calculated from it will differ slightly from run to run. Consequently, all results reported here are ensemble averaged over 1000 runs.

In order to determine the activation function and pinning current of the BSN, we first calculate the quantity $\langle \vec{s} \bullet \vec{m} \rangle$ as a function of the current amplitude $I_0$ after steady state is reached. Here $\vec{s}$ is the unit vector along the direction of the spin polarization in the injected current. When the spin-polarization is along the y-direction, $\langle \vec{s} \bullet \vec{m} \rangle = \langle m_y \rangle$. Similarly, when the spin-polarization is along the x-direction, $\langle \vec{s} \bullet \vec{m} \rangle = \langle m_x \rangle$. In Figs. 2(a) and 2(b), we show plots of $m_x(t)$ vs $t$ and $m_y(t)$ vs. $t$ for an elliptical nanomagnet (major axis = 100 nm, minor axis = 90 nm and thickness = 2 nm) for $I_0$ = 2 mA. The spin polarization in the injected current is in the +x direction and +y direction, respectively, for the two cases. The initial orientation of the magnetization is assumed to be always perpendicular to the axis of the current's spin polarization. Note that the time to achieve pinning in the two cases are different with the latter being longer. This is because the major axis (y-axis) is the easy axis and the minor axis (x-axis) is the hard axis. Since it is easier to pin the magnetization along the easy axis and harder to pin it along the hard axis, the time to pin is shorter in the former case.

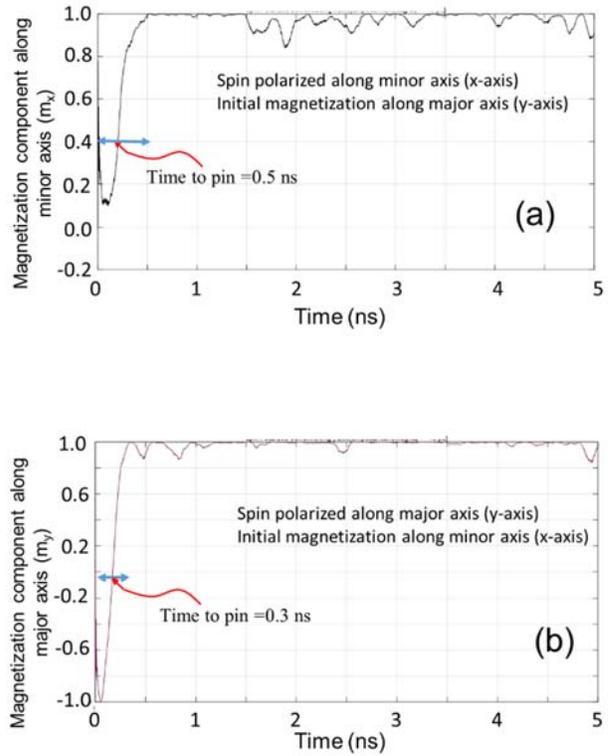

**Fig. 2**: Plots of (a) $m_x(t)$ vs $t$ and (b) $m_y(t)$ vs. $t$ for an elliptical nanomagnet (major axis = 100 nm, minor axis = 90 nm and thickness = 2 nm) for $I_0$ = 2.0 mA. The spin polarization in the injected current is in the +x direction and +y direction, respectively, for the two cases. The time to pin is the time taken to pin the magnetization in the orientation of the spin polarization. It has no relation to the response time or correlation time. The nutation time $\tau$ was taken to be 10 ps.

The time it takes for $m_x(t), m_y(t), m_z(t)$ to reach steady-state depends on the magnitude of $I_0$. Steady state is never reached if $I_0 = 0$. For the non-zero magnitudes of $I_0$ that we considered, steady state is always reached within 25 ns of turning on the current. Hence, we take the value of $m_x(t)$ or $m_y(t)$ at the end of 25 ns for every value of $I_0$ that we consider and then average that over 1000 runs to calculate $\langle \vec{s} \bullet \vec{m} \rangle = \langle m_x \rangle$ and $\langle \vec{s} \bullet \vec{m} \rangle = \langle m_y \rangle$ for that value of $I_0$. At steady-state, the system is ergodic and hence the time-average is equal to the ensemble-average.

In Fig. 3, we show the results of $\langle \vec{s} \bullet \vec{m} \rangle$ versus $I_0$ (i.e. the *activation function*) for three different values of the nutation duration τ associated with spin inertia, and find no significant τ-dependence. We consider three cases: (a) elliptical nanomagnet with the injected current spin-polarized along the major axis, (b) same nanomagnet with the injected current spin-polarized along the minor axis, and (c) perfectly circular nanomagnet. In all cases, the magnetization initially points perpendicular to the axis of spin polarization of the injected current (i.e., perpendicular to the intended direction of pinning). The positive and negative values of the current represent opposite (mutually antiparallel) spin polarizations. The curve has the expected $\tanh(I_0/I_c)$ [$I_c$ = constant] behavior [5, 6]. The two values of $I_0$ where $\langle \vec{s} \bullet \vec{m} \rangle$ reaches the value ±1 are the two *pinning currents* for pinning the magnetization along the direction of spin polarization and making the probability of bit 1 either 0% or 100%. These results reveal that spin inertia does not affect the activation function or the pinning current perceptibly, which is good happenstance since spin inertia cannot be easily engineered.

However, there is clearly some dependence on whether the nanomagnet is circular or elliptical and whether the spin polarization of the injected current is along the major axis or minor axis in the case of the elliptical nanomagnet. In fact, a 10% variation in a principal axis dimension (minor axis varying from 100 nm to 90 nm while the major axis remains invariant at 100 nm, i, e. circular to slightly elliptical) decreases the pinning current from 1.0 mA to 0.5 mA when the current is spin-polarized along the major (easy) axis and increases the pinning current from 1.0 mA to 2.0 mA when the current is spin-polarized along the minor (hard) axis. This happens because it is easier to pin the magnetization along the easy axis and harder to pin it along the hard axis, whereas the circular nanomagnet has no distinction between easy and hard axes. Therefore, we expect that the pinning current will be least in an elliptical nanomagnet with spin polarization along the major axis and most in the same nanomagnet with the spin polarization along the minor axis, with the circular nanomagnet being in between. That is exactly what we observe.

There are two lessons learned from this study. The first is that slight shape changes (going from circular to slightly elliptical) can change the activation function and the pinning current significantly. This sensitivity can lead to large device to device variation which is deleterious for large-scale integration. The second is that the minimum pinning current for pinning the probability of bit 1 to either 0% or 100% (and hence minimum energy dissipation) is obtained if we use slightly elliptical nanomagnets and encode the probabilistic p-bit in the component of the magnetization along the *major axis*. Thus, the optimum choice for minimum energy dissipation is a slightly elliptical nanomagnet with the p-bit encoded in the magnetization component along the major axis.

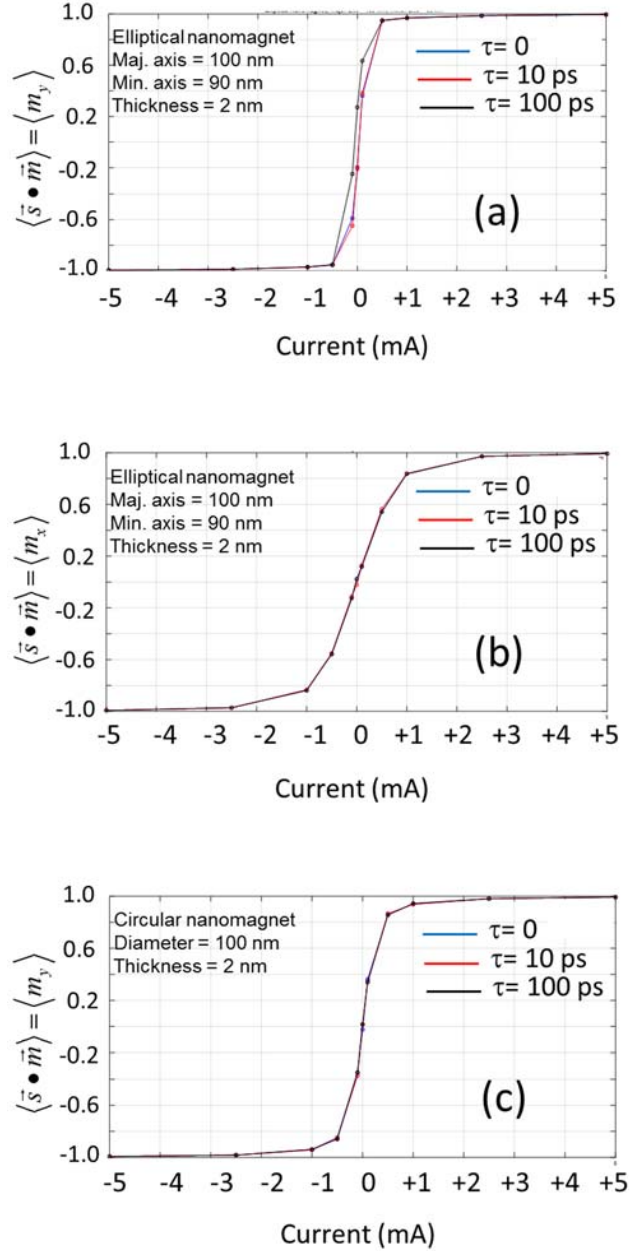

**Fig. 3**: Plot of the steady-state value of $\langle \vec{s} \bullet \vec{m} \rangle$ as a function of current for different values of the parameter τ representing the duration of nutation caused by spin inertia. (a) Elliptical nanomagnet (major axis = 100 nm, minor axis = 90 nm and thickness = 2nm) with an in-plane shape anisotropy energy barrier of 4.6 kT at room temperature and the current is spin-polarized along the *y*-axis or major axis (easy axis) (b) same nanomagnet when the current is spin-polarized along the *x*-axis or minor axis (hard axis), and (c) circular nanomagnet of diameter 100 nm and thickness 2 nm with no in-plane shape anisotropy energy barrier. The initial orientation of the magnetization is always orthogonal to the direction of spin polarization of the current.

### 3.2 *Response time of the BSN*

Next, we calculate the response time of the BSN. For this purpose, we first assume that the p-bit is encoded in the magnetization component along the minor axis, which we call the *x*-axis, and compute its auto-correlation function defined as $C_x(t') = \int_0^\infty m_x(t) m_x(t+t') dt$ in the absence of any spin-polarized current, starting with two initial conditions: the magnetization initially pointing approximately along the +*x* -axis and along the +*y*-axis. These two initial conditions are, respectively, $m_x(0) = 0.995; m_y(0) = 0.095; m_z(0) = 0.031$ and $m_x(0) = 0.095; m_y(0) = 0.995; m_z(0) = 0.031$. Starting with either initial condition, we follow the time trajectory of the magnetization component $m_x(t)$ [obtained by solving Equation (2)] to obtain the auto-correlation function $C_x(t')$ versus *t'* as $m_x(t)$ fluctuates between -1 and + 1 owing to thermal noise. Because the fluctuation history is slightly different from one run to another owing to the randomness of thermal noise, we average $C_x(t')$ versus *t'* over 1000 runs to obtain the final $C_x(t')$ versus *t'* plots.

The purpose of this calculation is to find the FWHM of the auto-correlation function, which is the correlation time $\tau_c$ that determines the response time of BSNs. In this case, the BSN should output bit 1 and bit -1 with equal probability (50% each) and we wish to estimate the time that will elapse before the BSN begins to output bits with this probability. This is the response time if the p-bit is encoded in the magnetization component parallel to the minor axis in an elliptical nanomagnet. Our purpose was to study if slight changes in the nanomagnet shape (from perfectly circular to slightly elliptical) can have a significant effect on the correlation time $\tau_c$ and hence the BSN response time.

We also obtain the auto-correlation function of the magnetization component along the *y*-axis which would be the major axis of the elliptical nanomagnet. This is defined as $C_y(t') = \int_0^\infty m_y(t) m_y(t+t') dt$. This will yield the response time if the p-bit is encoded in the magnetization component parallel to the major axis in an elliptical nanomagnet.

In Fig. 4, we plot the calculated auto-correlation functions as a function of the delay [i.e. $C_x(t')$ versus *t'* and $C_y(t')$ versus *t'*] for a slightly elliptical nanomagnet (major axis = 100 nm, minor axis = 99 nm) and a perfectly circular nanomagnet (diameter = 100 nm), both with the same thickness of 6 nm. The first has an in-plane shape anisotropy barrier of 4 *kT* at room temperature and the second has no such barrier. In the case of the slightly elliptical nanomagnet, we plot the auto-correlation functions for two cases: initial orientation of the magnetization is along the minor axis (*x*-axis) and initial orientation is along the major axis (*y*-axis).

In Table I, we list the calculated full-width-at-half-maximum (FWHM) of the auto-correlation functions (or the correlation time $\tau_c$) for the five cases. These times determine the response times of the corresponding BSNs.

From Table I, we find that spin inertia (i. e. the nutation time $\tau$) does not cause any significant change in the correlation time, but even a slight change in geometry (1% change in the diameter of the circular nanomagnet in any arbitrary direction to make it slightly elliptical)) can *decrease* the correlation time associated with fluctuation of the magnetization component along the minor axis by a factor of nearly 4 and *increase* the correlation time associated with fluctuation of the magnetization component along the major axis by a factor of nearly 4 as well. The opposite behaviors of the two magnetization components of course have to do with the fact that the major axis is the easy axis and the minor axis is the hard axis. The magnetization prefers to loiter around the easy axis and shun the hard axis. That is why the auto-correlation function of the fluctuation of the magnetization component along the easy axis decays much more slowly than that along the hard axis. What is surprising is that a very slight ellipticity (major axis = 100 nm and minor axis = 99 nm) can make so much difference. When probabilistic (p-) bits are encoded in the randomly fluctuating magnetization component of a slightly elliptical LBM [5-7], it makes a very significant difference as to whether the component along the major axis or the minor axis is chosen to encode the p-bit, since they have very different correlation times (and hence response times) that can differ by more than an order of magnitude.

TABLE I

THE CORRELATION TIME $\tau_c$ IN CIRCULAR AND SLIGHTLY ELLIPTICAL LBMS FOR THREE DIFFERENT SPIN INERTIA RELAXATION TIMES $\tau$

|  | $\tau$ = 0 ps | $\tau$ = 10 ps | $\tau$ = 100 ps |
|---|---|---|---|
| Circular LBM | 5.8 ns | 5.6 ns | 5.7 ns |
| $m_x$ in slightly elliptical nanomagnet. Initial orientation along major axis | 1.5 ns | 1.5 ns | 1.7 ns |
| $m_x$ in slightly elliptical nanomagnet. Initial orientation along minor axis | 1.6 ns | 1.6 ns | 1.7 ns |
| $m_y$ in slightly elliptical nanomagnet. Initial orientation along major axis | 20.25 ns | 21.07 ns | 20.24 ns |
| $m_y$ in slightly elliptical nanomagnet. Initial orientation along minor axis | 20.79 ns | 20.42 ns | 20.79 ns |

Earlier, we posited that if we wish to reduce energy dissipation, then we should encode the p-bit in the magnetization component along the *major* axis of a slightly elliptical LBM. However, that also makes the response much slower. There is an unavoidable trade-off between energy dissipation and speed in most information processing systems and that trend holds for this system as well.

We also notice that the initial magnetization orientation (whether it is pointing along the major axis or the minor axis) does not make any significant difference in the case of the slightly elliptical nanomagnet. One might have expected to see some dependence on the initial orientation because the major axis (*y*-axis) is the "easy axis" and hence corresponds to a stable direction while the minor axis (*x*-axis) is the "hard axis" and corresponds to an unstable direction. Therefore, if the magnetization is initially along the easy axis, we will expect it to resist straying from it while if it is initially along the hard axis, it should quickly leave that state. We do not see this effect since the decay time of the auto-correlation function exceeds 1 ns and over that long duration, the magnetization loses any memory of the initial orientation. That is why we see no effect of the initial state on the auto-correlation function decay characteristic or the correlation time $\tau_c$. Because the decay is relatively slow and occurs over a time scale exceeding 1 ns, which is *much longer* than the nutation time τ, we also do not see any effect of spin inertia. Sometimes spin inertia can have a very significant effect in slow magneto-dynamics that lasts over time scales orders of magnitude longer than τ [12], but that does not happen here.

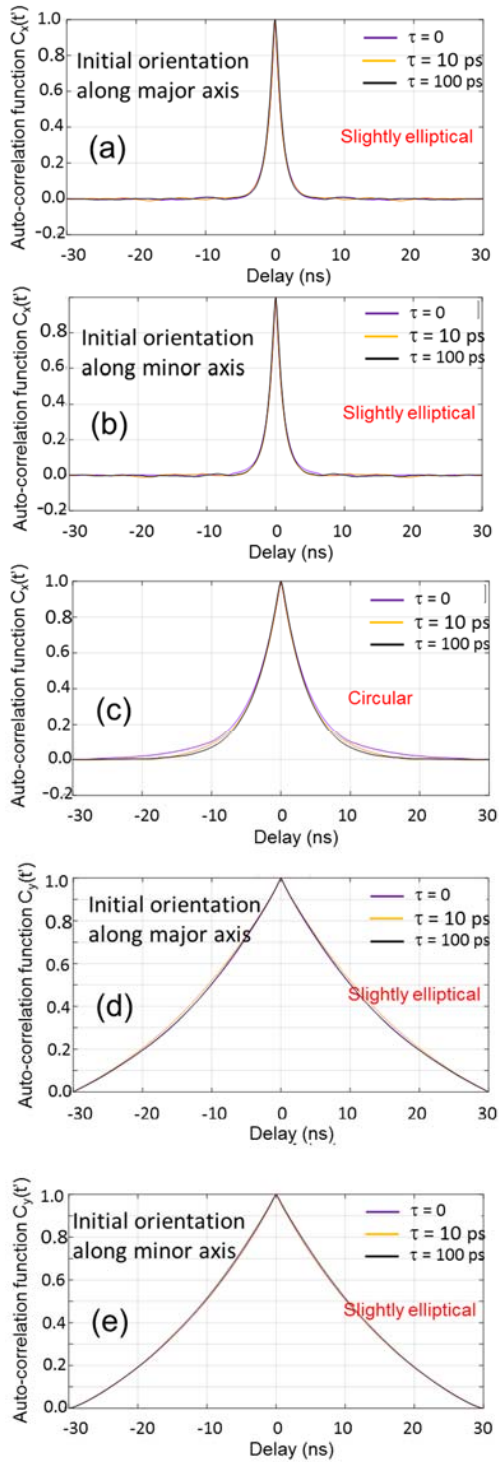

**Fig. 4**: Auto-correlation function of the fluctuations in the magnetization component along the *x*-axis (minor axis of the elliptical nanomagnet) versus delay shown for three different nutation times τ for (a) a slightly elliptical (100 nm, 99 nm) nanomagnet with initial orientation of the magnetization along the major axis, (b) the same slightly elliptical nanomagnet with initial orientation along the minor axis, and (c) a perfectly circular nanomagnet (diameter = 100 nm). Auto-correlation function of the fluctuations in the magnetization component along the *y*-axis (major axis) of the slightly elliptical nanomagnet versus delay with (d) initial orientation along the major axis and (e) initial orientation along the minor axis.

The most interesting point to note is that the correlation time in the circular nanomagnet differs by a factor of ~4 from that in the slightly elliptical nanomagnet (either 4× larger or 4× smaller depending on which magnetization component – along the major axis or along the minor axis – is chosen to encode the p-bit). The difference between the two nanomagnets is that the circular nanomagnet has no in-plane shape anisotropy energy barrier, whereas in the slightly elliptical nanomagnet, the in-plane shape anisotropy energy barrier is small (4 kT) but non-zero. Surprisingly, this small difference has a strong effect on the correlation time and hence the response speed of the BSN.

## IV. CONCLUSION

Low barrier nanomagnets (LBM) with in-plane anisotropy are a natural fit for BSNs, but unfortunately it turns out that their activation function, response time and pinning current are extremely sensitive to small geometric variations in their cross-section, which poses a serious problem for large-scale integration [17]. One possible way to mitigate this may be to introduce redundancy, i.e. encode a logical p-bit in multiple physical p-bits in the spirit of well-known quantum error correction schemes for qubits, but only at the cost of increased resources. A simpler approach has also been suggested [18]. We note that device-to-device variability is known to be a serious problem in memristor based neural networks [19, 20] which has prompted research in high entropy materials to counter this issue in memristors [21].

The activation function of a BSN plays a central role in the execution of algorithms, while the autocorrelation time $\tau_c$ determines how long it takes for a BSN based algorithm to produce a new output in response to a new input [22]. In Appendix II, we shown that device-to-device variation in the activation function of a restricted Boltzmann machine can cause significant variability in the root mean square error in training. If the autocorrelation time becomes uncertain and varies significantly from device to device, the execution of algorithms with BSNs is affected because we would not know how long to sample the BSN states to get the correct result. If we terminate the algorithm prematurely, then an incorrect result will be produced, as was demonstrated dramatically in ref. [23] which used stochastic nanomagnets to implement inverse AND logic. Finally, any uncertainty in the pinning current will introduce a mismatch in the drive currents, similar to the problem caused by threshold variability in MOSFET circuits, which can be debilitating in high-performance circuits [24]. In order to mitigate that problem, one would be forced to use excessive currents and hence suffer excessive energy dissipation.

In the past, we showed that extended structural defects in LBMs (e.g. random thickness variations) introduce a similar variability in the response times ($\tau_c$) of BSNs [25]. Here, we show that even slight variation in the lateral geometry can have a serious deleterious effect. In future, we will extend this to study the effect of edge roughness, which could be equally or more harmful.

We also show that the energy dissipation can be reduced by choosing slightly elliptical LBMs and encoding the p-bit in the magnetization component along the major axis, but this also increases the response time. Conversely, the response time can be reduced by encoding the p-bit in the magnetization component along the minor axis, but that increases the energy dissipation.

We can view the energy-delay product as being proportional to the product of the response time and the square of the pinning current for pinning the probability of either bit to 100%. For the cases considered here, the energy delay product turns out to be relatively independent of whether we use a circular or a slightly elliptical nanomagnet. In the elliptical case, the energy-delay product is also relatively independent of which magnetization component – the one along the major axis or the one along the minor axis – is chosen to encode the p-bit.

## APPENDIX I

Ref. [6] derived approximate analytical expressions for the correlation time and pinning currents. These expressions are

$$\tau_c = \sqrt{8\ln(2)}\frac{1}{\gamma}\sqrt{\frac{M_s\Omega}{4\pi\left|H_z^{\text{demag}}\right|kT}} \qquad (A1)$$

$$I_p = \frac{2q}{\hbar}\sqrt{\frac{2}{\pi}}\sqrt{4\pi\left|H_z^{\text{demag}}\right|M_s\Omega kT} \qquad ,(A2)$$

where $q$ is the electronic charge, $\hbar$ is the reduced Planck constant and $H_z^{\text{demag}}$ is the normal-to-plane demagnetizing field given by $H_z^{\text{demag}} = -M_s N_{d-zz}$. The other quantities were previously defined. These expressions were probably derived for the perfectly circular nanomagnet since no account is made of any shape anisotropy. It is informative to compare the analytical results with the numerically calculated results.

In our case, the quantity $H_z^{\text{demag}}$ is $7.25\times10^5$ A/m for the circular nanomagnet. The volume of the perfectly circular nanomagnet that we considered is 47,124 nm$^3$. Using these results in Equation (A1), we find that the correlation time $\tau_c$ is 333 ns in the perfectly circular nanomagnet, which is about two orders of magnitude larger than what we found and report in Table I.

The value of the pinning current calculated from Equation (A2) is about 2.9 A, which is three orders of magnitude larger than what we find in the circular nanomagnet. The expressions in Equations (A1) and (A2) are therefore overly pessimistic and predict much slower response and much higher current (hence much higher energy dissipation) than what is actually the case. It is not surprising that the analytical expressions are at odds with the numerical simulations because stochastic dynamics is rarely amenable to precise analytical treatment.

## APPENDIX II

To highlight the effect of variability in the activation function on training a neural network, we used an open source simulator for a restricted Boltzmann machine [26] where we redefined the sigmoid activation function as $sigmoid(z) = \tanh(\eta z)$ and varied $\eta$ between 0.8 and 1.0 to simulate the variability in the activation function. The results for the root mean square error (RMSE) in training versus number of iterations are shown in Fig. A2 for various values of $\eta$. One can see that there is significant variability in this plot if the activation function is varied.

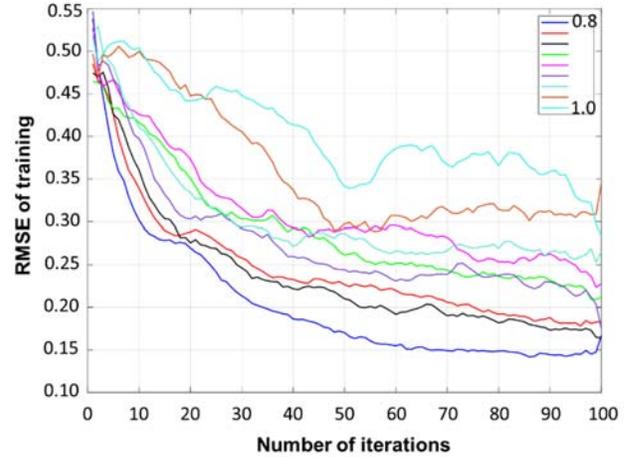

**Fig. A2**: Root mean square error in training a restricted Boltzmann machine of ref. [26] as a function of the number of iterations when the activation function $sigmoid(z) = \tanh(\eta z)$ is varied by varying $\eta$ from 0.8 to 1.0 in steps of 0.025. This illustrates the effect of device-to-device variations of the activation function on training a restricted Boltzmann machine and how sensitive the training is to device-to-device variations of the activation function.